\documentclass[runningheads]{llncs}

\usepackage{hyperref}
\usepackage{graphicx}

\usepackage{pgf,pgfrcs}
\usepackage{alg}

\begin{document}

\title{A Fixed Point Iteration Technique for Proving Correctness of Slicing for Probabilistic Programs
}

\titlerunning{A Fixed Point Iteration Technique}
\author{Torben Amtoft\inst{1}
\and 
Anindya Banerjee\inst{2}
}
\authorrunning{T. Amtoft and A. Banerjee}
\institute{Kansas State University, Manhattan KS 66506, USA
\email{tamtoft@ksu.edu}\\
\and
IMDEA Software Institute, Pozuelo de Alarcon, Madrid, Spain
\email{anindya.banerjee@imdea.org}}
\maketitle             

\begin{abstract}
When proving the correctness of a method for
slicing probabilistic programs,
it was previously discovered by the authors
that for a fixed point iteration to work
one needs a non-standard starting point for the iteration. 

This paper presents and explores this technique in a general setting;
it states the lemmas that must be established to
use the technique to prove the correctness of a program transformation,
and sketches how to apply the technique to slicing of probabilistic programs.


\end{abstract}

\newcommand{\Univ}{\ensuremath{\mathcal{U}}}

\newcommand{\stotwo}[4]{\ensuremath{\{{#1} \mapsto {#2},\; {#3} \mapsto {#4}\}}}
\newcommand{\stothree}[6]{\ensuremath{\{{#1} \mapsto {#2},\; {#3} \mapsto {#4},\; {#5} \mapsto {#6}\}}}
\newcommand{\stofour}[8]{\ensuremath{\{{#1} \mapsto {#2},\; {#3} \mapsto {#4},\; {#5} \mapsto {#6},\; {#7} \mapsto {#8}\}}}

\newcommand{\lesseq}{\sqsubseteq}
\newcommand{\least}{\bot}
\newcommand{\chain}[2]{\{{#1} \mid {#2}\}}
\newcommand{\arrow}{\rightarrow}
\newcommand{\contarrow}{\rightarrow_{c}}
\newcommand{\lub}[1]{\sqcup {#1}}
\newcommand{\fixed}[1]{\mathit{fix}({#1})}
\newcommand{\rel}[3]{{#1}\;{#2}\;{#3}}
\newcommand{\startN}{\mathsf{start}}
\newcommand{\stopN}{\mathsf{end}}
\newcommand{\Dist}{\ensuremath{{\sf D}}}

\section{Setting}
We wish to reason about the correctness of a program transformation
of a ``source semantics'' $\phi$ into a ``target semantics'' $\gamma$,
each semantics defined on a universe $\Univ$ of ``subprograms''.
Typically, such correctness is established by demonstrating
some ``correctness relation'' $R$ such that $\rel{R}{\phi}{\gamma}$
can be proved.
But we shall show that for a certain application,
the standard technique for such a proof
cannot be adapted directly but requires modification.

We assume that for each $x \in \Univ$,
$\phi(x)$ and $\gamma(x)$ are \emph{continuous}
functions from $\Dist$ to $\Dist$
where $\Dist$ is a \emph{complete partial order} (cpo).
Recall\footnote{Throughout we employ elementary concepts
of domain theory; see, e.g., Winskel's 
textbook~\cite{Winskel1993}.} that
\begin{itemize}
\item
a cpo $\Dist$ is a set equipped
with a partial order $\lesseq$ that satisfies:
\begin{quote}
if $\chain{d_i}{i \geq 0}$ is a chain
(that is $d_i \lesseq d_{i+1}$ for all $i \geq 0$)
then it has a least upper bound, written $\lub{\chain{d_i}{i \geq 0}}$,
thus:
\begin{itemize}
\item
$d_i \lesseq \lub{\chain{d_i}{i \geq 0}}$ for all $i \geq 0$, and
\item
if $d_i \lesseq d'$ for all $i \geq 0$
then $\lub{\chain{d_i}{i \geq 0}} \lesseq d'$.
\end{itemize}
\end{quote}
If in addition there is a least element
$\least$ (thus $\least \lesseq d$ for all $d \in \Dist$)
we say that $\Dist$ is a \emph{pointed} cpo.
\item
a function $f$, from a cpo $\Dist_1$ into a cpo $\Dist_2$,
is continuous if for all chains
$\chain{d_i}{i \geq 0}$ in $\Dist_1$ it holds that
\begin{itemize}
\item
$\chain{f(d_i)}{i \geq 0}$ is a chain in $\Dist_2$ 
(this implies that $f$ is monotone)
\item
$f (\lub{\chain{d_i}{i \geq 0}}) =
\lub{\chain{f(d_i)}{i \geq 0}}$.
\end{itemize}
We write $\Dist_1 \contarrow \Dist_2$ for the space of continuous
functions from $\Dist_1$ to $\Dist_2$.
\item
if $\Dist_1$ and $\Dist_2$ are cpos then also
$\Dist_1 \contarrow \Dist_2$ is a cpo, with the partial
order defined pointwise, as is the least upper bound:
\[
\lub{\chain{f_k}{k \geq 0}} = \lambda d \in \Dist_1. \lub{\chain{f_k(d)}{k \geq 0}}
\]
and if $\least$ is least in $\Dist_2$ then
$\lambda d.\least$ is least in $\Dist_1 \contarrow \Dist_2$.
\end{itemize}

\section{Goal}
We want $\phi$ and $\gamma$ to be related as prescribed by
a given correctness relation
$R$ over $(\Dist \contarrow \Dist)$;
that is we shall require $\rel{R}{\phi}{\gamma}$
which is an abbreviation of
\begin{equation}
\label{eq:goal}
\forall x \in \Univ: \rel{R}{(\phi\; x)}{(\gamma\; x)}.
\end{equation}
We shall aim at finding a widely applicable recipe for
establishing (\ref{eq:goal}), but shall see that the 
most straightforward
approach does not always work.

We shall assume that $R$ is \emph{admissible} in that
if $\chain{f_i}{i \geq 0}$ and
$\chain{g_i}{i \geq 0}$ are chains in $\Dist \contarrow \Dist$ then
\begin{equation}
\label{eq:admissible}
\mbox{if } 
\rel{R}{f_i}{g_i} \mbox{ for all } i \geq 0 \mbox{ then }
\rel{R}{(\lub{\chain{f_i}{i \geq 0}})}{(\lub{\chain{g_i}{i \geq 0}})}.
\end{equation}

\section{Example Setting}
We shall consider \emph{probabilistic programming}
(with \cite{Kozen:JCSS-81} a seminal contribution
and \cite{Gor+etal:ICSE-2014} giving a recent overview)
where the effect of a program is to transform probability distributions,
and shall express a program as a 
\emph{probabilistic control-flow graph}~\cite{Amtoft+Banerjee:TOPLAS-2020}.
In that setting,
\begin{itemize}
\item
the subprograms in $\Univ$ are of the form
$(u,u')$ where $u,u'$ are nodes such that $u'$ postdominates $u$
(that is, $u'$ occurs on every path from $u$ to an end node).
\item
the members of $\Dist$ are probability distributions
where a probability distribution $D$
maps a store,
which is a (partial) map from program variables
to integers or booleans, into a real number in $[0..1]$ which is 
the ``probability'' of that store.

With $D_1 \lesseq D_2$ defined to hold iff 
$D_1(s) \leq D_2(s)$ for all stores $s$,
$\Dist$ will be a pointed cpo where
\begin{itemize}
\item the least element is $\lambda s.0$
\item the least upper bound of a chain
$\chain{D_i}{i \geq 0}$ is $\lambda s.S(s)$
where $S(s)$ is the supremum of the chain 
$\chain{D_i(s)}{i \geq 0}$.
\end{itemize}
\end{itemize}
We can think of
$\phi(u,u')$ as denoting the distribution transformation done 
along the path(s) from $u$ to $u'$: if 
$D$ describes how stores are distributed at point $u$,
then $\phi(u,u')(D)$ describes how stores are distributed at point $u'$.

\section{Example}
\begin{figure}[p]
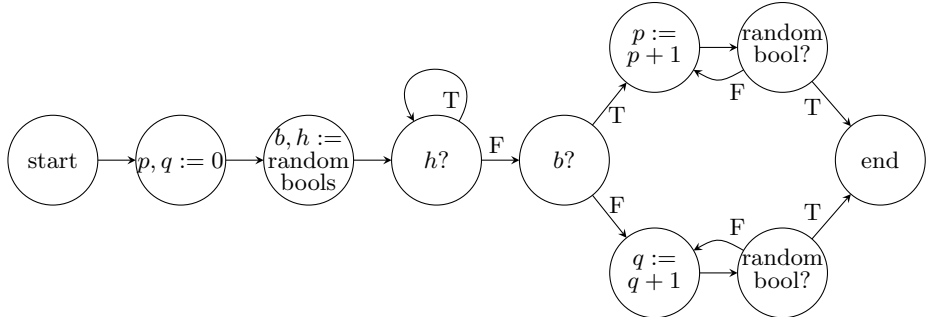


\begin{pgfpicture}{10mm}{23mm}{120mm}{57mm}

\pgfnodecircle{s}[stroke]{\pgfxy(2,4)}{17pt}
\pgfputat{\pgfxy(2,3.9)}{\pgfbox[center,base]{start}}

\pgfnodecircle{pq}[stroke]{\pgfxy(3.7,4)}{17pt}
\pgfputat{\pgfxy(3.7,3.9)}{\pgfbox[center,base]{$p,q := 0$}}

\pgfnodecircle{bh}[stroke]{\pgfxy(5.4,4)}{17pt}
\pgfputat{\pgfxy(5.4,4.2)}{\pgfbox[center,base]{$b,h :=$}}
\pgfputat{\pgfxy(5.4,3.9)}{\pgfbox[center,base]{random}}
\pgfputat{\pgfxy(5.4,3.6)}{\pgfbox[center,base]{bools}}

\pgfnodecircle{h}[stroke]{\pgfxy(7.1,4)}{17pt}
\pgfputat{\pgfxy(7.1,3.9)}{\pgfbox[center,base]{$h$?}}

\pgfnodecircle{b}[stroke]{\pgfxy(8.8,4)}{17pt}
\pgfputat{\pgfxy(8.8,3.9)}{\pgfbox[center,base]{$b$?}}

\pgfnodecircle{p1}[stroke]{\pgfxy(10,5.5)}{17pt}
\pgfputat{\pgfxy(10,5.6)}{\pgfbox[center,base]{$p := $}}
\pgfputat{\pgfxy(10,5.3)}{\pgfbox[center,base]{$ p+1$}}

\pgfnodecircle{q1}[stroke]{\pgfxy(10,2.5)}{17pt}
\pgfputat{\pgfxy(10,2.6)}{\pgfbox[center,base]{$q := $}}
\pgfputat{\pgfxy(10,2.3)}{\pgfbox[center,base]{$ q+1$}}

\pgfnodecircle{rp}[stroke]{\pgfxy(11.7,5.5)}{17pt}
\pgfputat{\pgfxy(11.7,5.6)}{\pgfbox[center,base]{random}}
\pgfputat{\pgfxy(11.7,5.3)}{\pgfbox[center,base]{bool?}}

\pgfnodecircle{rq}[stroke]{\pgfxy(11.7,2.5)}{17pt}
\pgfputat{\pgfxy(11.7,2.6)}{\pgfbox[center,base]{random}}
\pgfputat{\pgfxy(11.7,2.3)}{\pgfbox[center,base]{bool?}}

\pgfnodecircle{e}[stroke]{\pgfxy(13,4)}{17pt}
\pgfputat{\pgfxy(13,3.9)}{\pgfbox[center,base]{end}}

\pgfsetarrowsend{stealth} 

\pgfnodeconnline{s}{pq}
\pgfnodeconnline{pq}{bh}
\pgfnodeconnline{bh}{h}
\pgfnodeconnline{h}{b}
\pgfnodeconnline{b}{p1}
\pgfnodeconnline{b}{q1}
\pgfnodeconnline{p1}{rp}
\pgfnodeconnline{q1}{rq}
\pgfnodeconnline{rp}{e}
\pgfnodeconnline{rq}{e}

\pgfnodeconncurve{h}{h}{60}{120}{30pt}{30pt}
\pgfnodeconncurve{rp}{p1}{210}{330}{10pt}{10pt}
\pgfnodeconncurve{rq}{q1}{150}{30}{10pt}{10pt}

\pgfputat{\pgfxy(12.1,4.6)}{\pgfbox[center,base]{T}}
\pgfputat{\pgfxy(12.1,3.2)}{\pgfbox[center,base]{T}}

\pgfputat{\pgfxy(11.1,4.8)}{\pgfbox[center,base]{F}}
\pgfputat{\pgfxy(11.1,3.0)}{\pgfbox[center,base]{F}}

\pgfputat{\pgfxy(7.9,4.1)}{\pgfbox[center,base]{F}}
\pgfputat{\pgfxy(7.3,4.7)}{\pgfbox[center,base]{T}}

\pgfputat{\pgfxy(9.5,4.5)}{\pgfbox[center,base]{T}}
\pgfputat{\pgfxy(9.5,3.3)}{\pgfbox[center,base]{F}}

\end{pgfpicture}

\caption{\label{fig:CFG} The source control-flow graph of our running example.}
\end{figure}

We shall consider the probabilistic structured program in 
Figure~\ref{fig:source}
which
\begin{enumerate}
\item
assigns 0 to the variables $p$ and $q$
\item
assigns random boolean values to 
the variables $b$ and $h$,
with the choices independent and unbiased
so that each outcome (say $b = \mbox{true}$ and $h = \mbox{false}$)
has probability $\frac{1}{2} \cdot \frac{1}{2} = \frac{1}{4}$ 
\item
if $h$ is true, enters an infinite loop 
\item
if $b$ is true (false), increments $p$ ($q$)
until a random boolean choice becomes true
(which in each iteration will happen with
probability $\frac{1}{2}$)
\end{enumerate}
The corresponding
control-flow graph is depicted in 
Figure~\ref{fig:CFG}.

\begin{figure}[p]
\begin{algorithm}
\begin{algtab*}
$p$ := 0; $q$ := 0 \\
$b$ := random boolean value \\
$h$ := random boolean value \\
\algif{$h$}
\algwhile{true}
skip \\
\algend
\algelse
skip \\ 
\algend
\algif{$b$}
\algrepeat 
$p$ := $p+1$ \\
\alguntil{random boolean choice becomes true}
\algelse
\algrepeat
$q$ := $q+1$ \\
\alguntil{random boolean choice becomes true}
\algend
\end{algtab*}
\end{algorithm}
\caption{\label{fig:source} The source code of our running example.}
\end{figure}

We will expect that for arbitrary initial $D$,
the distribution $\phi(\startN,\stopN)\,(D)$ 
is given by Table~\ref{table:source}.
\begin{table}[p]
\[
\begin{array}{c|cl}
 s    & \phi(\startN,\stopN)\,(D)  & \\ \hline
\stofour{b}{T}{h}{F}{p}{k}{q}{0} & \displaystyle \frac{1}{2^{k+2}}
& \mbox{ for } k \geq 1 \\[2mm]
\stofour{b}{F}{h}{F}{p}{0}{q}{k} & \displaystyle \frac{1}{2^{k+2}}
& \mbox{ for } k \geq 1 
\end{array}
\]
\caption{\label{table:source} The distribution produced by
the source semantics.}
\end{table}
To understand why say
$\stofour{b}{F}{h}{F}{p}{0}{q}{2}$ has probability 
$\frac{1}{2^{2+2}} = \frac{1}{2^4}$,
observe that to obtain this store we need four fortunate binary outcomes:
\begin{enumerate}
\item
$b$ becomes $F$ (so that $q$, rather than $p$, is incremented)
\item
$h$ becomes $F$ (so the infinite loop is not taken)
\item
the first random choice is $F$ (so $q$ will become 2)
\item
the second random choice is $T$ (so $q$ remains 2)
\end{enumerate}
Observe that the probabilities add up to $\frac{1}{2}$, since if $h = T$
we never get to $\stopN$.

\section{Slicing the Example}
In general, if we care about only certain variables,
we would like to \emph{slice} away
all parts of the program that do not affect those variables;
see Danicic et al.~\cite{Danicic+etal:TCS-2011}
for a unifying framework for program slicing.

For our example, assume that the part of the store
we care about is the value of $q$, whereas we do not
care about the value of $p$.
Observe that we still need to care about the value of $b$ 
because it is \emph{relevant} in that
it determines whether $q$ is incremented,
but $p$ is irrelevant,
as is $h$ as it does not impact the \emph{relative} distribution of $q$
(since $h$ and $b$ are independent).
Accordingly, we can abstract
the distribution from Table~\ref{table:source}
into Table~\ref{table:source1}.

\begin{figure}[p]
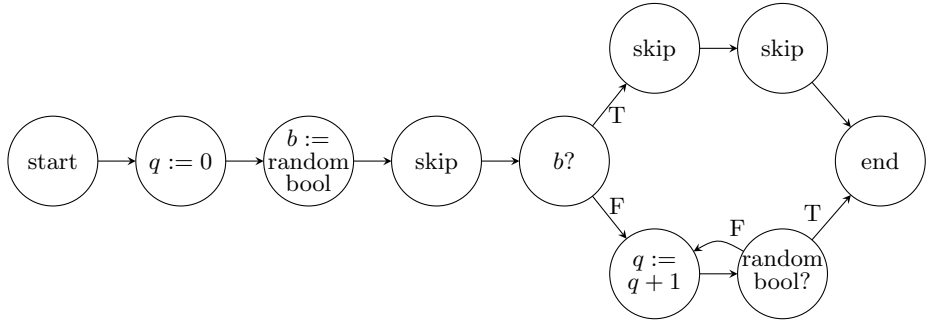


\begin{pgfpicture}{10mm}{23mm}{120mm}{57mm}

\pgfnodecircle{s}[stroke]{\pgfxy(2,4)}{17pt}
\pgfputat{\pgfxy(2,3.9)}{\pgfbox[center,base]{start}}

\pgfnodecircle{pq}[stroke]{\pgfxy(3.7,4)}{17pt}
\pgfputat{\pgfxy(3.7,3.9)}{\pgfbox[center,base]{$q := 0$}}

\pgfnodecircle{bh}[stroke]{\pgfxy(5.4,4)}{17pt}
\pgfputat{\pgfxy(5.4,4.2)}{\pgfbox[center,base]{$b :=$}}
\pgfputat{\pgfxy(5.4,3.9)}{\pgfbox[center,base]{random}}
\pgfputat{\pgfxy(5.4,3.6)}{\pgfbox[center,base]{bool}}

\pgfnodecircle{h}[stroke]{\pgfxy(7.1,4)}{17pt}
\pgfputat{\pgfxy(7.1,3.9)}{\pgfbox[center,base]{skip}}

\pgfnodecircle{b}[stroke]{\pgfxy(8.8,4)}{17pt}
\pgfputat{\pgfxy(8.8,3.9)}{\pgfbox[center,base]{$b$?}}

\pgfnodecircle{p1}[stroke]{\pgfxy(10,5.5)}{17pt}
\pgfputat{\pgfxy(10,5.4)}{\pgfbox[center,base]{skip}}

\pgfnodecircle{q1}[stroke]{\pgfxy(10,2.5)}{17pt}
\pgfputat{\pgfxy(10,2.6)}{\pgfbox[center,base]{$q := $}}
\pgfputat{\pgfxy(10,2.3)}{\pgfbox[center,base]{$ q+1$}}

\pgfnodecircle{rp}[stroke]{\pgfxy(11.7,5.5)}{17pt}
\pgfputat{\pgfxy(11.7,5.4)}{\pgfbox[center,base]{skip}}

\pgfnodecircle{rq}[stroke]{\pgfxy(11.7,2.5)}{17pt}
\pgfputat{\pgfxy(11.7,2.6)}{\pgfbox[center,base]{random}}
\pgfputat{\pgfxy(11.7,2.3)}{\pgfbox[center,base]{bool?}}

\pgfnodecircle{e}[stroke]{\pgfxy(13,4)}{17pt}
\pgfputat{\pgfxy(13,3.9)}{\pgfbox[center,base]{end}}

\pgfsetarrowsend{stealth} 

\pgfnodeconnline{s}{pq}
\pgfnodeconnline{pq}{bh}
\pgfnodeconnline{bh}{h}
\pgfnodeconnline{h}{b}
\pgfnodeconnline{b}{p1}
\pgfnodeconnline{b}{q1}
\pgfnodeconnline{p1}{rp}
\pgfnodeconnline{q1}{rq}
\pgfnodeconnline{rp}{e}
\pgfnodeconnline{rq}{e}

\pgfnodeconncurve{rq}{q1}{150}{30}{10pt}{10pt}

\pgfputat{\pgfxy(12.1,3.2)}{\pgfbox[center,base]{T}}

\pgfputat{\pgfxy(11.1,3.0)}{\pgfbox[center,base]{F}}

\pgfputat{\pgfxy(9.5,4.5)}{\pgfbox[center,base]{T}}
\pgfputat{\pgfxy(9.5,3.3)}{\pgfbox[center,base]{F}}

\end{pgfpicture}

\caption{\label{fig:CFGtarget} The target control-flow graph of our example program.}
\end{figure}

We then slice the given program so as to eliminate
irrelevant operations, such as the assignments to $p$ and $h$,
and even the loop on $h$ 
(as we consider \emph{termination-insensitive} slicing).
Figure~\ref{fig:CFGtarget} shows the resulting
control-flow graph, which can be further simplified
and corresponds to the structured program in Figure~\ref{fig:target}.

That control-flow graph induces the target semantics $\gamma$.
It is easy to see that for arbitrary initial $D$,
the distribution $\gamma(\startN,\stopN)\,(D)$
is given by Table~\ref{table:target}.

\begin{table}[p]
\[
\begin{array}{c|cl}
 s    & \phi(\startN,\stopN)\,(D)\; s  & \\ \hline
\stotwo{b}{T}{q}{0} & \displaystyle \frac{1}{4} &
\\[2mm]
\stotwo{b}{F}{q}{k} & \displaystyle \frac{1}{2^{k+2}}
& \mbox{ for } k \geq 1 
\end{array}
\]
\caption{\label{table:source1} The relevant part of the
distribution produced by
the source semantics.}
\end{table}

\begin{figure}[p]
\begin{algorithm}
\begin{algtab*}
$q$ := 0; $b$ := random boolean value \\
\algif{$b$}
skip \\
\algelse
\algrepeat
$q$ := $q+1$ \\
\alguntil{random boolean choice becomes true}
\algend
\end{algtab*}
\end{algorithm}
\caption{\label{fig:target} The target code of our example program.}
\end{figure}

\begin{table}[p]
\[
\begin{array}{c|cl}
 s     & \gamma(\startN,\stopN)\,(D)\,(s)  & \\ \hline 
\stotwo{b}{T}{q}{0} &
 \displaystyle \frac{1}{2}
& \\[2mm]
\stotwo{b}{F}{q}{k} & \displaystyle \frac{1}{2^{k+1}}
& \mbox{ for } k \geq 1 
\end{array}
\]
\caption{\label{table:target} The distribution produced by
the target semantics.}
\end{table}
By comparing Tables~\ref{table:source1} and~\ref{table:target}
we see:
\[
   \phi(\startN,\stopN)\,(D) = \frac{1}{2} \cdot
\gamma(\startN,\stopN)\,(D) 
\]
and indeed, for this application, our correctness relation $R$
requires:
\[
R\; (\phi(u,u'))\; (\gamma(u,u')) 
\mbox{ iff } \forall D \in \Dist\; \exists c :
\phi(u,u')(D) = c \cdot \gamma(u,u')(D).
\]
That is, the transformation must not affect the
\emph{relative} probabilities of the various stores.

We shall now study how to \emph{prove} that this correctness relation holds.

\section{Basic Approach}

Recall the fixed-point theorem for cpos:
\begin{quote}
if $\Dist$ is a pointed cpo, and $F$ belongs to $\Dist \contarrow \Dist$,
then 
$\chain{F^i(\least)}{i \geq 0}$ is a chain with
\[ 
F(\lub{\chain{F^i(\least)}{i \geq 0}}) = \lub{\chain{F^i(\least)}{i \geq 0}}.
\]
The chain property follows by induction, with the base case being
$F^0(\least) = \least \lesseq F^1(\least)$,
and the inductive step using that $F$ is monotone:
\[
F^{i+1}(\least) = F(F^i(\least)) \lesseq F(F^{i+1}(\least))
=
F^{i+2}(\least).
\]
That the least upper bound is a fixed point follows
from the continuity of $F$:
\[
F(\lub{\chain{F^i(\least)}{i \geq 0}}) = 
\lub{\chain{F(F^i(\least))}{i \geq 0}}  
= \lub{\chain{F^i(\least)}{i \geq 0}}
\]
and it is actually the least fixed point since if also $f$
is a fixed point then for all $i \geq 0$
we have $F^i(\least) \lesseq f$, as can be proved by induction
with the inductive step being $F^{i+1}(\least) = F(F^i(\least))
\lesseq F(f) = f$.
\end{quote}
It will typically be the case that
$\phi$ and $\gamma$ have been defined as
\begin{eqnarray*}
\phi    & = & \lub{\chain{F^i(\least)}{i \geq 0}} \\
\gamma  & = & \lub{\chain{G^i(\least)}{i \geq 0}} 
\end{eqnarray*}
for functionals $F$ and $G$ with 
functionality
\[
 F,G : (\Univ \arrow \Dist \contarrow \Dist) \contarrow (\Univ \arrow \Dist \contarrow \Dist).
\]
where 
$\Univ \arrow \Dist \contarrow \Dist$ is a pointed cpo since
\begin{itemize}
\item
$\Dist \contarrow \Dist$ is a pointed cpo, since $\Dist$ is
\item
equipped with the discrete partial order,
$\Univ$ is a cpo and 
all functions from $\Univ$ are continuous.
\end{itemize}
Thus the fixed point theorem can be applied;
we shall often write $\phi_k$ for $F^k(\least)$ and
$\gamma_k$ for $G^k(\least)$.

To accomplish our goal (\ref{eq:goal}),
that is to prove that $\rel{R}{\phi}{\gamma}$,
due to $R$ being admissible (\ref{eq:admissible})
it thus suffices to show that
\begin{equation}
\label{eq:correctK}
\rel{R}{\phi_k}{\gamma_k} \mbox{ for all } k \geq 0
\end{equation}
which indeed often can be proved by induction in $k$.
But we shall now see that (\ref{eq:correctK}) may \emph{not}
always hold.

\section{Problem}
In the setting of control flow graphs,
we will expect $\phi_k$ and $\gamma_k$ to be the meaning of 
the program assuming we are allowed at most $k-1$
``backwards moves'', that is control flowing away from the end node
(as will necessarily happen each time a loop is iterated).
In particular, $\phi_1$ will allow
only forward moves, and thus we would expect 
(ignoring the values of $h$ and $p$) that
$\phi_1(\startN,\stopN)\,(D)$ is given by
\[
\begin{array}{c|cl}
 s    & \phi_1(\startN,\stopN)\,(D)  & \\ \hline
\stotwo{b}{T}{q}{0} & \displaystyle \frac{1}{8}
& 
\\ [2mm]
\stotwo{b}{F}{q}{1} & \displaystyle \frac{1}{8}
\end{array}
\]
whereas $\phi_2$ allows for at most one backwards move and
thus we would expect that
$\phi_2(\startN,\stopN)\,(D)$ is given by
\[
\begin{array}{c|cl}
 s    & \phi_2(\startN,\stopN)\,(D)  & \\ \hline
\stotwo{b}{T}{q}{0} & \displaystyle \frac{1}{8} + \frac{1}{16}
& 
\\ [2mm]
\stotwo{b}{F}{q}{1} & \displaystyle \frac{1}{8}
\\ [2mm]
\stotwo{b}{F}{q}{2} & \displaystyle \frac{1}{16}
\end{array}
\]
Thus the probability that execution ends with $q = 0$
is $\frac{3}{16}$, which makes sense since for this to 
happen we need two independent events:
\begin{itemize}
\item
$b$ becomes $T$ and $h$ becomes $F$; this has probability $\frac{1}{4}$
\item
the random choice at the upper branch becomes $T$
either first time or second time; this has probability $\frac{3}{4}$.
\end{itemize}
In general, for $k \geq 1$,
we would expect that the distribution 
$\phi_k(\startN,\stopN)\,(D)$ is given by
Table~\ref{table:sourcek},
in particular we have 
\begin{eqnarray*}
& & \phi_k(\startN,\stopN)\,(D)\;\stotwo{b}{T}{q}{0}
\\ & = & \displaystyle  \sum_{i = 1}^k 
\phi_k(\startN,\stopN)\,(D)\;\stothree{b}{T}{p}{i}{q}{0}
\\
& = & \displaystyle \sum_{i = 1}^k 
\frac{1}{2^{i+2}}
= \frac{1}{8} + \frac{1}{16} + \ldots \frac{1}{2^{k+2}}
= \frac{1}{4} - \frac{1}{2^{k+2}}
\end{eqnarray*}

\begin{table}[p]
\[
\begin{array}{c|cl}
 s    & \phi_k(\startN,\stopN)\,(D)  & \\ \hline
\stotwo{b}{T}{q}{0} & \displaystyle \frac{1}{4} - \frac{1}{2^{k+2}}
& 
\\ [2mm]
\stotwo{b}{F}{q}{i} & \displaystyle \frac{1}{2^{i+2}}
& \mbox{ when } 1 \leq i \leq k
\end{array}
\]
\caption{\label{table:sourcek} The relevant part of
the source distribution after 
$k \geq 1$ iterations.}
\end{table}

On the other hand, looking at the target control-flow graph
(Figure~\ref{fig:CFGtarget}),
we see that no backwards moves are needed 
for $q$ to end up being 0, and hence 
we would expect that the distribution 
$\gamma_k(\startN,\stopN)\,(D)$ is given by
Table~\ref{table:targetk}.

\begin{table}[p]
\[
\begin{array}{c|cl}
 s    & \gamma_k(\startN,\stopN)\,(D)  & \\ \hline
\stotwo{b}{T}{q}{0} & \displaystyle \frac{1}{2}
& 
\\ [2mm]
\stotwo{b}{F}{q}{i} & \displaystyle \frac{1}{2^{i+1}}
& \mbox{ when } 1 \leq i \leq k
\end{array}
\]
\caption{\label{table:targetk} The relevant part of
the target distribution after 
$k \geq 1$ iterations.}
\end{table}
By comparing Tables~\ref{table:sourcek} and~\ref{table:targetk}
we see that 
(\ref{eq:correctK}) does {\bf not} hold, in particular
\[
   \phi_k(\startN,\stopN)\,(D) \neq \frac{1}{2} \cdot
\gamma_k(\startN,\stopN)\,(D).
\]
We shall now show how to repair this,
by replacing $\chain{\phi_k}{k \geq 0}$
by a more appropriate chain.

\section{Refined Approach}
To prepare for the correctness proof, the idea is
to first transform the source control-flow graph.
This involves
replacing loops that
are irrelevant, and eventually terminate, by {\sf skip}.

\begin{figure}[p]
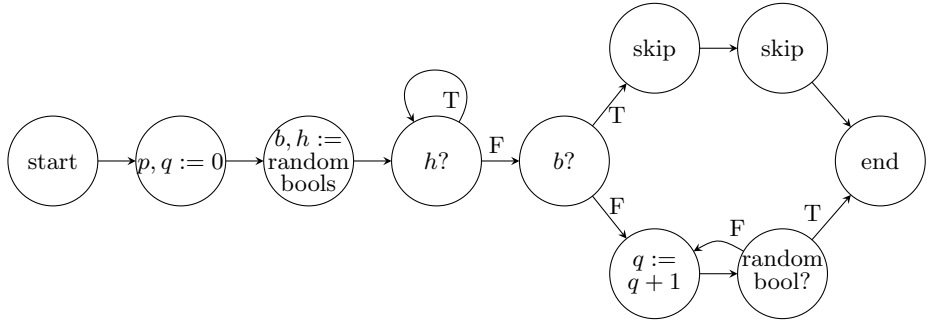


\begin{pgfpicture}{10mm}{23mm}{120mm}{57mm}

\pgfnodecircle{s}[stroke]{\pgfxy(2,4)}{17pt}
\pgfputat{\pgfxy(2,3.9)}{\pgfbox[center,base]{start}}

\pgfnodecircle{pq}[stroke]{\pgfxy(3.7,4)}{17pt}
\pgfputat{\pgfxy(3.7,3.9)}{\pgfbox[center,base]{$p,q := 0$}}

\pgfnodecircle{bh}[stroke]{\pgfxy(5.4,4)}{17pt}
\pgfputat{\pgfxy(5.4,4.2)}{\pgfbox[center,base]{$b,h :=$}}
\pgfputat{\pgfxy(5.4,3.9)}{\pgfbox[center,base]{random}}
\pgfputat{\pgfxy(5.4,3.6)}{\pgfbox[center,base]{bools}}

\pgfnodecircle{h}[stroke]{\pgfxy(7.1,4)}{17pt}
\pgfputat{\pgfxy(7.1,3.9)}{\pgfbox[center,base]{$h$?}}

\pgfnodecircle{b}[stroke]{\pgfxy(8.8,4)}{17pt}
\pgfputat{\pgfxy(8.8,3.9)}{\pgfbox[center,base]{$b$?}}

\pgfnodecircle{p1}[stroke]{\pgfxy(10,5.5)}{17pt}
\pgfputat{\pgfxy(10,5.4)}{\pgfbox[center,base]{skip}}

\pgfnodecircle{q1}[stroke]{\pgfxy(10,2.5)}{17pt}
\pgfputat{\pgfxy(10,2.6)}{\pgfbox[center,base]{$q := $}}
\pgfputat{\pgfxy(10,2.3)}{\pgfbox[center,base]{$ q+1$}}

\pgfnodecircle{rp}[stroke]{\pgfxy(11.7,5.5)}{17pt}
\pgfputat{\pgfxy(11.7,5.4)}{\pgfbox[center,base]{skip}}

\pgfnodecircle{rq}[stroke]{\pgfxy(11.7,2.5)}{17pt}
\pgfputat{\pgfxy(11.7,2.6)}{\pgfbox[center,base]{random}}
\pgfputat{\pgfxy(11.7,2.3)}{\pgfbox[center,base]{bool?}}

\pgfnodecircle{e}[stroke]{\pgfxy(13,4)}{17pt}
\pgfputat{\pgfxy(13,3.9)}{\pgfbox[center,base]{end}}

\pgfsetarrowsend{stealth} 

\pgfnodeconnline{s}{pq}
\pgfnodeconnline{pq}{bh}
\pgfnodeconnline{bh}{h}
\pgfnodeconnline{h}{b}
\pgfnodeconnline{b}{p1}
\pgfnodeconnline{b}{q1}
\pgfnodeconnline{p1}{rp}
\pgfnodeconnline{q1}{rq}
\pgfnodeconnline{rp}{e}
\pgfnodeconnline{rq}{e}

\pgfnodeconncurve{h}{h}{60}{120}{30pt}{30pt}
\pgfnodeconncurve{rq}{q1}{150}{30}{10pt}{10pt}

\pgfputat{\pgfxy(12.1,3.2)}{\pgfbox[center,base]{T}}

\pgfputat{\pgfxy(11.1,3.0)}{\pgfbox[center,base]{F}}

\pgfputat{\pgfxy(7.9,4.1)}{\pgfbox[center,base]{F}}
\pgfputat{\pgfxy(7.3,4.7)}{\pgfbox[center,base]{T}}

\pgfputat{\pgfxy(9.5,4.5)}{\pgfbox[center,base]{T}}
\pgfputat{\pgfxy(9.5,3.3)}{\pgfbox[center,base]{F}}

\end{pgfpicture}

\caption{\label{fig:CFGprep} 
The modified source control-flow graph of our example program.}
\end{figure}

In our setting, with $p$ considered irrelevant,
this will result in the control-flow graph depicted in 
Figure~\ref{fig:CFGprep}.

With that control-flow graph as our starting point,
we may define a chain $\chain{\phi'_k}{k \geq 0}$
where for $k \geq 1$,
we would expect that the distribution 
\newline
$\phi'_k(\startN,\stopN)\,(D)$ is given by
Table~\ref{table:sourcekk}.

\begin{table}[p]
\[
\begin{array}{c|cl}
 s    & \phi'_k(\startN,\stopN)\,(D)  & \\ \hline
\stotwo{b}{T}{q}{0} & \displaystyle \frac{1}{4}
& 
\\ [2mm]
\stotwo{b}{F}{q}{i} & \displaystyle \frac{1}{2^{i+2}}
& \mbox{ when } 1 \leq i \leq k
\end{array}
\]
\caption{\label{table:sourcekk} The relevant part of
the revised source distribution after 
$k \geq 1$ iterations.}
\end{table}

By comparing Tables~\ref{table:sourcekk} and~\ref{table:targetk}
we see that for $k \geq 1$,
\[
   \phi'_k(\startN,\stopN)\,(D) = \frac{1}{2} \cdot
\gamma_k(\startN,\stopN)\,(D) 
\]
which suggests that we do indeed have
$\rel{R}{\phi'_k}{\gamma_k}$ when $k \geq 1$.

This is a special case,
of the general idea that we shall now present.

\section{General Result}
\label{sec:result}
Recall the setting:
with $\Dist$ a pointed cpo, $\Univ$ a set of subprograms,
$F$ and $G$ functionals of type
$(\Univ \arrow \Dist \contarrow \Dist) \contarrow (\Univ \arrow \Dist \contarrow \Dist)$,
and with $R$ an admissible
correctness relation, we want to prove
$\rel{R}{\phi}{\gamma}$ where
\begin{eqnarray*}
\phi & = & \lub{\chain{F^{i}(\least)}{i \geq 0}} 
 \\[1mm]
\gamma & = & \lub{\chain{G^{i}(\least)}{i \geq 0}}. 
\end{eqnarray*}
For that purpose, we might be able to use the following result:
\begin{theorem}
\label{thm:main}
With $X \subseteq \Univ$ given, 
for $k \geq 0$ define $\phi'_k = F^k(\phi'_0)$
with $\phi'_0$ given by
\[
\begin{array}{rcll}
  \phi'_0(x) & = & \phi(x) & \mbox{ if } x \in X \\
  \phi'_0(x) & = & \least & \mbox{ otherwise }
\end{array}
\]
(thus the starting point for the iteration
may be \emph{above} the standard starting point $\least$),
and similarly for $k \geq 0$ define $\gamma'_k = G^k(\gamma'_0)$
where $\gamma'_0$ is given by:
\[
\begin{array}{rcll}
  \gamma'_0(x) & = & \gamma(x) & \mbox{ if } x \in X \\
  \gamma'_0(x) & = & \least & \mbox{ otherwise. }
\end{array}
\]
Now assume that the below 3 properties all hold:
\begin{equation}
\label{iterate}
\forall k \geq 0 : \rel{R}{\phi'_k}{\gamma'_k}
\end{equation}
\begin{equation}
\label{fixf}
\forall x \in X : \phi(x) \lesseq \phi'_1(x)
\end{equation}
\begin{equation}
\label{fixg}
\forall x \in X : \gamma(x) \lesseq \gamma'_1(x).
\end{equation}
Then we do have 
$\rel{R}{\phi}{\gamma}$.
\end{theorem}
\begin{proof}
Observe that due to (\ref{iterate}) and (\ref{eq:admissible}),
our goal $\rel{R}{\phi}{\gamma}$
will follow if we can prove that
\begin{quote}
$\chain{\phi'_k}{k \geq 0}$ is a chain with
$\lub{\chain{\phi'_k}{k \geq 0}} = \phi$ 
\\[2mm]
$\chain{\gamma'_k}{k \geq 0}$ is a chain with
$\lub{\chain{\gamma'_k}{k \geq 0}} = \gamma$ 
\end{quote}
and by symmetry it suffices to prove the first property. To do so,
first observe that
\[
\forall x \in \Univ : \phi'_0(x) \lesseq \phi'_1(x)
\]
which is trivial if $x \notin X$
and otherwise follows from (\ref{fixf}).
By monotonicity of $F$, this implies that
$\chain{\phi'_k}{k \geq 0}$ is a chain.
Writing $\phi_i$ for $F^i(\least)$ (with $i \geq 0$),
we obviously have
\[
\phi_0 \lesseq \phi'_0 \lesseq \phi
\]
which implies (since $F$ is monotone and has $\phi$ as a fixed point)
\[
\phi_1 \lesseq \phi'_1 \lesseq F(\phi) = \phi
\]
and in general (by induction in $k$)
\[
\phi_k \lesseq \phi'_k \lesseq \phi \mbox{ for all } k \geq 0
\]
Because $\lub{\chain{\phi_k}{k \geq 0}} = \phi$, we obtain the desired
$\lub{\chain{\phi'_k}{k \geq 0}} = \phi$.
\end{proof}
To summarize, we see that the successful
application of our approach requires choosing
a suitable set $X$, to be called an \emph{exclusion set}
as it is excluded from the iteration process, and 
then establishing (\ref{iterate}), (\ref{fixf}), (\ref{fixg}).

In the next section, we shall see that
our example setting does allow a suitable $X$ to be found;
we shall not address whether there are heuristics for
that in a general setting.

\section{An Application}
We have presented a recipe
for proving the correctness of an algorithm for slicing
probabilistic programs. An elaborate, albeit unmotivated, proof of correctness
was furnished
in the authors' earlier work~\cite{Amtoft+Banerjee:TOPLAS-2020}.
In contrast, the technique proposed in Section~\ref{sec:result}
abstracts the essence of the proof structure,
and clarifies the key lemmas that are needed.

In~\cite{Amtoft+Banerjee:TOPLAS-2020}, slicing involves
finding suitable node sets $Q$ and $Q_0$,
where $Q$ are those nodes that impact the variable(s) of interest,
whereas $Q_0$ are those of the remaining nodes that may
cause non-termination.
The exclusion set $X$ (cf.~Section~\ref{sec:result}) is then
defined to contain the pairs $(v,v') \in \Univ$ 
where (using the terminology of \cite{Amtoft+Banerjee:TOPLAS-2020})
``$v$ stays outside $Q \cup Q_0$ until $v'$'',
a property which 
(as proved in \cite{Amtoft+Banerjee:TOPLAS-2020}[Lemma 5.8]) ensures
that from $v$, control will eventually (without risk of non-termination)
reach $v'$, without affecting relevant variables.

For the control-flow graph in
Figure~\ref{fig:CFG}, two nodes are not in $Q \cup Q_0$:
the node that increments $p$, and its successor
(the node that tests $h$ may cause non-termination, and all other
nodes impact $q$).
With $u_p$ the node that increments $p$ we thus have
\begin{equation}
\label{upX}
(u_p,\stopN) \in X
\end{equation}
In Section~\ref{sec:result} we listed the properties
sufficient for correctness:
(\ref{iterate}), (\ref{fixf}), (\ref{fixg})
and admissibility (\ref{eq:admissible}).
They are all stated
in~\cite{Amtoft+Banerjee:TOPLAS-2020} (though 
with quite different naming 
conventions\footnote{\cite{Amtoft+Banerjee:TOPLAS-2020} uses
``$\omega$'' for what the current paper calls ``$\phi$'',
uses ``$\gamma$'' for ``$\phi'$'', uses ``$\phi$'' for ``$\gamma$'',
uses ``$\Phi$'' for ``$\gamma'$'', uses ``$\mathsf{H}_V$'' for ``$F$'',
and uses ``$\mathsf{H}_Q$'' for ``$G$''.}):
\begin{itemize}
\item
property (\ref{iterate}) is stated as 
\cite{Amtoft+Banerjee:TOPLAS-2020}[Lemma 6.9]
(proved by induction in $k$, and with a case analysis
on $(v,v') \in \Univ$)
\item
property (\ref{fixf}) is a special case of 
\cite{Amtoft+Banerjee:TOPLAS-2020}[Lemma 6.2]
which (renamed) states that
 $\phi'_k(x) = \phi(x)$ for all $x \in X$ 
and all $k \geq 0$
\item
property (\ref{fixg}) is a special case of 
\cite{Amtoft+Banerjee:TOPLAS-2020}[Lemma 4.31]
which (renamed) entails 
that $G(g_1)(x) = G(g_2)(x)$ for all $x \in X$
and all functions $g_1,g_2$,
from which we for $x \in X$
infer $\gamma(x) = G(\gamma)(x) = G(\gamma'_0)(x)
= \gamma'_1(x)$
\item
property (\ref{eq:admissible}) follows from the calculation in
\cite{Amtoft+Banerjee:TOPLAS-2020}[Proof of Theorem 6.6],
except that we need to \emph{rectify} 
\cite{Amtoft+Banerjee:TOPLAS-2020}[Lemma 6.13].
That lemma claims
that a certain sequence of reals is a chain;
this is not necessarily the case,
but 
(as shown in Appendix~\ref{sec:rectify})
still the sequence will have a limit, and that is sufficient
to establish property (\ref{eq:admissible}).
\end{itemize}

\section{The Intuition Behind the Exclusion Set}
In our application, where the source control-flow graph of
Figure~\ref{fig:CFG} is transformed into
the target control-flow graph of Figure~\ref{fig:CFGtarget}.
consider paths from $\startN$ to $\stopN$ along which
$b$ becomes true, thus causing 
the value of $q$ at $\stopN$ to be 0.
\begin{itemize}
\item
For the target program (Figure~\ref{fig:CFGtarget}),
the only such path is the one that takes the upper branch.
That path has no loops and 
hence the fixed point
(which assigns $q = 0$ the probability $\frac{1}{2}$)
is reached already in the first iteration,
as can be seen by comparing the first row of 
Table~\ref{table:targetk} with the first row of Table~\ref{table:target}.
\item
On the other hand, for the source program (Figure~\ref{fig:CFG}), 
there are denumerable many such paths; all will follow the upper branch
but one path will not loop, another path will loop once, another
path will loop twice, etc.
Hence the fixed point 
(which assigns $q = 0$ the probability $\frac{1}{4}$)
is reached only in the limit of the iterations,
as can be seen by comparing the first row of 
Table~\ref{table:sourcek} with the first row of Table~\ref{table:source1}.
\end{itemize}
We conclude that source iteration and target iteration get out of lockstep,
which is why we cannot establish the correctness relation between
the iterands.

Our approach is to let the iteration on the source program 
start with the parts in the exclusion set $X$ already at their fixed point.
This is accomplished by ensuring that if $(u,u')$ in $X$
then no path from $u$ to $u'$ has loops.
Recall (\ref{upX}) that with $u_p$ the node that increments $p$
we have $(u_p,\stopN) \in X$.
And indeed, no path from $u_p$ to $\stopN$ has loops
in the control-flow graph in 
Figure~\ref{fig:CFGprep}, as 
we put {\sf skip} at node $u_p$ and at its successor. 

As a consequence,
the revised iterands of the source program (Table~\ref{table:sourcekk})
are in lockstep with the iterands of the target program 
(Table~\ref{table:targetk}),
with each iterand of the source being $\frac{1}{2}$ of the
corresponding iterand of the target.

\section{Perspectives}
We have distilled a proof technique 
that in at least one (recently published) situation 
\cite{Amtoft+Banerjee:TOPLAS-2020}
comes to the rescue when
the standard technique falls short.
We conjecture that other applications exist,
and encourage the discovery of such.

More generally, while we were surprised that we had
to begin the fixed point iteration
above the bottom element, we believe it likely that
other researchers have encountered and reported 
similar situations,
though so far we have not come across it in
our literature search.

We offer this paper as a tribute to Alan Mycroft
who has inspired the authors since early in their careers,
by his kindness and enthusiasm, and his work on program
analysis where the computation of fixed points
is ubiquitous.
\medskip

\textbf{Acknowledgments.} We thank the anonymous reviewers for their comments which in various ways helped improve the presentation of the paper, and also thank Patrick Cousot for looking into our findings. Banerjee’s research is based on work supported by the National Science Foundation (NSF), while working at the Foundation. Any opinions, findings, and conclusions or recommendations expressed in this paper are those of the authors and do not necessarily reflect the views of the NSF.

\bibliographystyle{splncs04}
\bibliography{main}

\appendix

\section{Rectifying Lemma 6.13 in \cite{Amtoft+Banerjee:TOPLAS-2020}}
\label{sec:rectify}

\newcommand{\limit}[2]{\mathit{lim}_{#1 \rightarrow \infty}\,{#2}}
\newcommand{\cnst}[4]{\ensuremath{c_{{#1},{#4}}^{{#2,#3}}}}

We shall refer to the setting of \cite{Amtoft+Banerjee:TOPLAS-2020},
where
$\Univ$ is the set of pairs $(v,v')$ such that $v'$ postdominates $v$,
and  $\Dist$ is the set of probability distributions.
With $\cnst{k}{v}{v'}{D}$ a real number as defined in Definition~6.10,
we shall replace Lemma 6.13
by the following result:
\begin{quote}
{\bf Lemma 6.13, restated}.
Assume $\chain{D_k}{k \geq 0}$ is a chain in $\Dist$,
and that $\chain{f_k}{k \geq 0}$ is a chain of functions
in $\Dist \contarrow \Dist$.

Then for all $(v,v') \in \Univ$, the limit
$\limit{k}{\cnst{k}{v}{v'}{f_k(D_k)}}$ does exist.
\end{quote}
Observe that as a special case, with each $f_k$ the identity
and each $D_k$ equal $D$, we have
\begin{quote}
For all $(v,v') \in \Univ$ and $D \in \Dist$,
there does exist a real number $c$ such that
\[
\limit{k}{\cnst{k}{v}{v'}{D}} = c
\]
\end{quote}
which suffices to prove Theorem 6.6,
stating the correctness of slicing in
\cite{Amtoft+Banerjee:TOPLAS-2020}.

We shall prove the restated Lemma 6.13 by induction
in the maximum length of an acyclic path
from $v$ to $v'$,  where two cases are non-trivial;
in each we let $v''$ denote the first proper postdominator of $v$:
\begin{itemize}
\item
Assume $v' \neq v$ and $v' \neq v''$.
We can inductively assume that there exists $c'', c'$ such that
\begin{eqnarray*}
\limit{k}{\cnst{k}{v}{v''}{f_k(D_k)}} & = & c''
\\
\limit{k}{\cnst{k}{v''}{v'}{\gamma_k^{(v,v'')}(f_k(D_k))}} & = & c'
\end{eqnarray*}
but then
\[
\limit{k}{\cnst{k}{v}{v'}{f_k(D_k)}} =
\limit{k}{(\cnst{k}{v}{v''}{f_k(D_k)} \cdot
\cnst{k}{v''}{v'}{\gamma_k^{(v,v'')}(f_k(D_k))})}
=
c'' \cdot c'.
\]
\item
Assume $v' = v''$ but $v$ does not stay outside $Q_0$ until $v'$.

Since $\chain{f_k(D_k)}{k \geq 0}$ and
$\chain{\gamma_k^{(v,v')}(f_k(D_k))}{k \geq 0}$ 
are chains in the cpo $\Dist$,
there exists $D_1$ and $D_2$ such that
\begin{eqnarray*}
\limit{k}{f_k(D_k)} & = & D_1
\\
\limit{k}{\gamma_k^{(v,v')}(f_k(D_k))} & = & D_2.
\end{eqnarray*}
By Lemma 4.1 we see that
\begin{eqnarray*}
\limit{k}{(\sum f_k(D_k))} & = & \sum D_1
\\
\limit{k}{(\sum \gamma_k^{(v,v')}(f_k(D_k)))} & = & \sum D_2.
\end{eqnarray*}
If $D_1 = 0$ we for all $k$ have
$f_k(D_k) = 0$ and thus
$\cnst{k}{v}{v'}{f_k(D_k)} = 1$ making the claim trivial.
We can thus assume that $D_1 \neq 0$,
in which case we see that
\[
\limit{k}{\cnst{k}{v}{v'}{f_k(D_k)}}
=
\limit{k}{\frac{\sum \gamma_k^{(v,v')}(f_k(D_k))}
{\sum f_k(D_k)}} = \frac{\sum D_2}{\sum D_1}
\]
which yields the claim.

But observe that while
$\chain{\cnst{k}{v}{v'}{f_k(D_k)}}{k \geq 0}$
has a limit, 
we apparently have no assurance it is a chain
(it is rather a \emph{ratio} between two chains)
and thus the original Lemma 6.13 makes an unwarranted claim.
\end{itemize}

\end{document}